# Meta-surface-enabled ultra-sharp multimode waveguide bends on silicon

Hao Wu[+1], Chenlei Li[+1], Lijia Song[1], Hon-Ki Tsang [2], John E Bowers [3], and Daoxin Dai*[1]

[1] Centre for Optical and Electromagnetic Research, State Key Laboratory for Modern Optical Instrumentation, College of Optical Science and Engineering, East Building No.5, Zhejiang University, Hangzhou 310058, China.

[2] Department of Electronic Engineering, the Chinese University of Hong Kong, Shatin, N. T., Hong Kong, S. A. R., China.

[3] Department of Electrical and Computer Engineering, University of California, Santa Barbara, CA 93106, USA.

## Abstract

Mode-division-multiplexing (MDM) is attractive as a means to increase the link capacity of a single wavelength for optical interconnects via the use of multiple mode-channels in multimode bus waveguides. In order to route the data in MDM systems, waveguide bends are typically needed. In particular, ultra-sharp bends are desirable for reducing the chip area occupied by high-density photonic integrated circuits (PICs). In this work, we propose and demonstrate a novel meta-surfaced waveguide structure on silicon, which enables ultra-sharp multimode waveguide bends to have low-loss and low-crosstalk among different mode-channels. An ultra-sharp S-bend with a bending radius of $R$=10μm is realized for an MDM optical interconnect with three mode-channels. The numerical simulation shows that the proposed ultra-sharp meta-surfaced multimode waveguide bend (MMWB) has low excess losses (0.2~0.5 dB) and low inter-mode crosstalk (<−30 dB) over a broad wavelength-band (>100 nm) for all the three mode-channels. This is verified by experimental results, which show that the fabricated S-bends have excess losses of 0.5~0.8dB and inter-mode crosstalk of <−20 dB over a wavelength-band of >60 nm. The experimental results were partially limited by the measurement setup, particularly the on-chip mode (de)multiplexers which had inter-mode crosstalk of about −20 dB. The proposed MMWB can be extended for the transmission of more than three mode-channels. Our work paves the way to use meta-surfaced silicon photonic waveguide structures for on-chip multimode optical interconnects.

+ These authors contributed equally to this work.
* Corresponding Author: dxdai@zju.edu.cn

Bandwidth requirements for data transmission is increasing exponentially for a wide range of optical networks, including on-chip optical interconnects [1], [2]. Further enhancement of the link capacity of optical interconnects has been a perennial goal of scientists and engineers globally [2]-[6]. A cost-effective solution is utilizing advanced multiplexing technologies [2],[5], including wavelength-division-multiplexing (WDM) [7], polarization-division-multiplexing (PDM) [8], mode-division-multiplexing (MDM) [9], and space-division-multiplexing (SDM) [10]. WDM and PDM have been widely deployed, and MDM is very attractive as an emerging technology because of the potential for significantly increasing the link capacity by utilizing multiple channels of guided modes [9]-[14]. The challenge is the effective manipulation of all the guided modes, including not only the fundamental mode but also the higher-order modes. There are two key elements for achieving low-loss and low-crosstalk MDM optical interconnects. *One* is a low-loss and low-crosstalk on-chip mode (de)multiplexer for efficiently combining/separating the signals carried by different mode-channels in MDM systems. Recently on-chip mode (de)multiplexers have been developed by utilizing various silicon photonic structures, including multimode interference (MMI) couplers [21],[22], adiabatic mode-evolution couplers [23], asymmetric Y-junctions [24]-[26], topology structures [27], and asymmetric directional couplers (ADCs) [28]-[31]. These reported devices employed innovative design and advanced fabrication to achieve low-loss (<1 dB) and low-crosstalk (~−20 dB) MDM for transmitters and receivers.

The other key element for MDM systems is a multimode bus waveguide with sharp bends enabling *low-loss* and *low-crosstalk* transmission for multiple mode-channels. Sharp bends are indispensable for multimode waveguide interconnects in photonic networks-on-chip to enable efficient use of valuable chip area. As is well known, for a *singlemode* optical waveguide, what is concerned is to achieve low bending loss only, and sharp bends can easily be achieved with low loss by introducing a high index-contrast $\Delta$. For example, for a *singlemode* silicon-on-insulator (SOI) nanophotonic waveguide, low-loss sharp bend with a micro-scale bending radius $R$ (e.g., ~3μm) can be achieved due to the ultra-high index-contrast $\Delta$ [32]. Sharp bends can be also realized even for a subwavelength-grating waveguide, which has reduced index-contrast $\Delta$, by introducing a pre-distorted refractive index profile to reduce the radiation loss and mode mismatch loss [33]. For a *singlemode* subwavelength-grating waveguide bend with $R$=5 μm, it was shown that the bending loss was reduced from 5.4 dB/90° to 1.1 dB/90° [33].

However, for a *multimode* SOI nanophotonic waveguide, the situation becomes very complicated because all modal fields become highly asymmetric when the bending radius becomes small [34]. As a consequence, there is significant mode mismatch between a straight waveguide (SWG) and a sharp bent waveguide (BWG), which introduces not only notable excess losses but also high inter-mode crosstalk [34]-[37]. As a result, one has to choose very large bending radii when using regular arc-bends. For example, for a 4μm-wide multimode bus waveguide, the bending radius $R$ for the regular arc-bend should be millimeter scale [35]. A possible approach to achieve a sharp bend for multimode bus waveguides is to use a specially-designed bent section whose curvature is modified from a small value (close to zero) to a given value $1/R_0$ gradually, in which way the guided-modes in the straight section can be converted gradually to the guided-modes in the bent section. This approach has already been used for realizing low-loss *singlemode* waveguide bends [32], and has also been suggested for realizing compact low-loss and low-crosstalk multimode waveguide bends [34]. However, this type of multimode waveguide bend is still as large as ~100 μm because it is required to make all the guided-modes convert adiabatically. Another solution is to introduce some special mode converters, which are designed to eliminate the mode-mismatch between the straight section and the bent section. In Ref. [36], [37], a step-tapered mode converter was demonstrated to realize a compact waveguide bend with $R$=5μm for the fundamental mode and the first higher-order mode. Unfortunately, it is difficult to be extended for the multimode waveguide bends with more higher-order modes. In 2012, Lipson et al. proposed a multimode waveguide bend with a modified cross section, designed using transformation optics [35]. With this design, the bending radius for a multimode waveguide bend with three guided-modes could be as compact as ~78 μm. However, the excess loss is high (i.e., ~2.5dB in theory, and ~2.6dB in experiment) and a special grayscale lithography process is needed for the fabrication.

In this paper, we propose and demonstrate a novel ultra-sharp multimode waveguide bend by introducing meta-surface structures to enable low-loss and low-crosstalk MDM on-chip optical interconnects. Silicon-based all-dielectric metasurfaces have been very attractive for potentially low loss and used widely for light at normal-incidence to the surface [38]. Recently Silicon-based metasurface structures consisting of phased arrays of plasmonic or dielectric nanoantennas were reported to realize waveguide mode converters and polarization rotators [39]. Here we introduce silicon meta-surface structure to spatially modify the refractive index profile in the bent section, so that the mode profiles in the bent section has very little mode mismatch with those of the straight multimode bus waveguide. In this way, the excess loss and the inter-mode crosstalk introduced by the bent section are very low even when the bending radius is ultra-small. Furthermore, the proposed multimode waveguide bend can be fabricated using standard complementary-metal-oxide-semiconductor (CMOS) processes for silicon photonics. As an example, a sharp multimode waveguide bend with $R = 10$ μm is demonstrated to support the transmission with three mode-channels (i.e., $M_{ch} = 3$). In theory this ultra-sharp multimode waveguide bend has a low excess loss (0.2~0.5 dB) and a low inter-mode crosstalk (<−30 dB) over a broad wavelength-band (>100 nm) for all the three mode-channels. The experimental result shows that the fabricated S-bend has excess losses of 0.5~0.8 dB and inter-mode crosstalk of less than −20 dB in a wavelength-range exceeding 60 nm. The measurements were limited by the integrated on-chip mode (de)multiplexers which had inter-mode crosstalk of ~−20 dB. The proposed MMWB can be extended for transmission of more than three mode-channels.

# Structure and Principle

Figure 1(a)-(b) show the schematic configuration of the proposed MMWB with a constant bending radius $R$ and a core width $w_{co}$. Two uniform SWGs with the same core widths are connected at the input/output ends of the MMWB. This MMWB is designed optimally to minimize the mode mismatch between SWGs and MMWB. For the present MMWB, a shallowly-etched non-uniform sub-wavelength grating is formed on the top surface of the silicon core, as shown in Figure 1(c). This meta-surface structure is defined with a center-period $p_0$ and a center-duty-cycle $\eta_0$ along the central axis, while the duty-cycle $\eta$ varies linearly in the radical direction. One has

$$\eta(\rho)= \eta_0+\gamma\rho, \tag{1}$$

where $\rho$ is the coordinate along the radial direction ($\rho$=0 at the central axis of the waveguide core), $\gamma$ is the asymmetry-constant determining the variant ratio of the duty-cycle in the radical direction. The duty-cycles at the two sidewalls of the MMWB are given as $\eta_1= \eta_0 + \gamma\, w_{co}/2$ and $\eta_2= \eta_0 − \gamma\, w_{co}/2$, respectively. It can be seen that a larger $|\gamma|$ indicates the meta-surface with more asymmetry. In order to compensate the asymmetry due to the bending, one should have $\gamma$<0 so that $\eta_2$<$\eta_0$<$\eta_1$. Furthermore, since $0 \leq \eta_2 < \eta_1 \leq 1$, the constant $\gamma$ should be chosen so that $0 \leq \gamma \leq \min[\eta_0/(w_{co}/2), (1-\eta_0)/(w_{co}/2)]$. As an example, one has $\gamma \geq 1/w_{co}$ when assuming $\eta_0=0.5$. The period is chosen as $p = 0.25$ μm to satisfy the requirement of supporting the Bloch-Floquet mode [40].

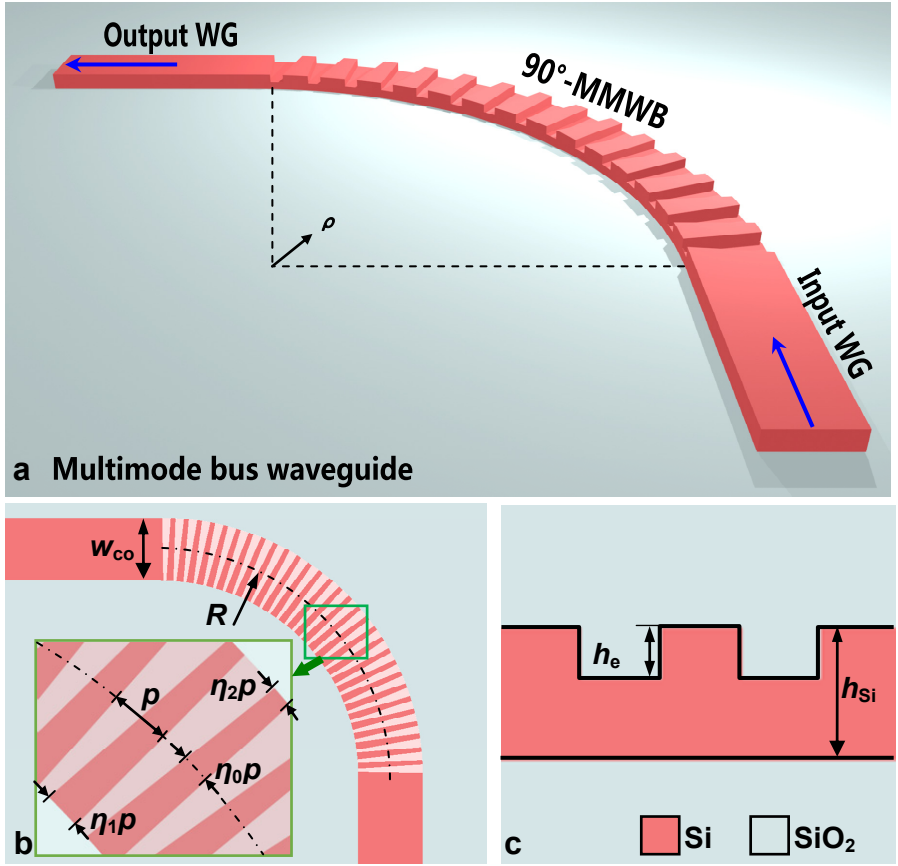

**Figure 1: Proposed meta-surfaced multimode waveguide bend (MMWB).** (a) Three-dimensional schematic configuration of the proposed 90°-MMWB; (b) Top-view configuration of the proposed MMWB (Inset: the enlarged view of the meta-surfaced structure); (c) Cross-section of the MMWB along the central axis.

# Results

In order to make the structural optimization simplified, we make the present MMWB equivalent to a dual-core BWG, as shown in Figure 2(a). According to the effective medium theory [41], here the meta-surfaced silicon layer at the top of the silicon core region is equivalent to an inhomogeneous layer with an effective index $n_{\mathrm{eff}}(\rho)$ (see Figure 2(b)). The effective index $n_{\mathrm{eff}}(\rho)$ is determined by the duty-cycle $\eta(\rho)$ of the meta-surfaced structure [42], i.e.,

$$n_{\mathrm{eff}}^2(\rho) = \eta(\rho) n_1^2 + [1-\eta(\rho)] n_2^2, \tag{2}$$

where $n_1$ and $n_2$ are the refractive indices of the waveguide core and cladding, respectively. In the present case, one has $n_1=3.48$ (silicon) and $n_2=1.444$ (silica), respectively [43]. For such an equivalent dual-core optical waveguide with any given $\gamma$, one can conveniently use the overlapping integral method to estimate the excess loss and the inter-mode crosstalk introduced by the mode mismatching, when any mode-channel is launched to propagate along the photonic circuit consisting of an input

SWG, an MMWB with a given bending radius $R$, and an output SWG (see Method).

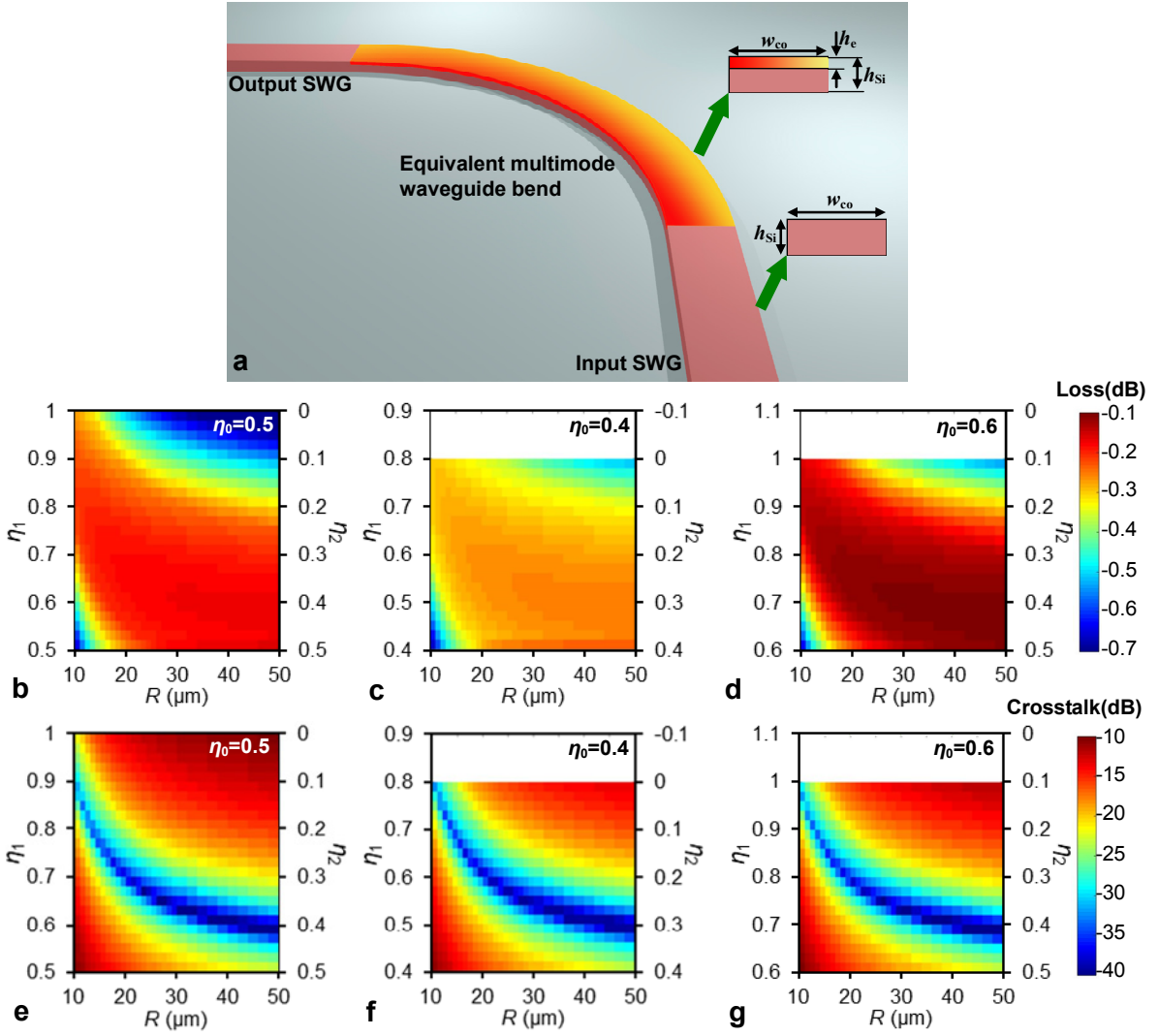

Figure 2: **Configuration and simulation for an equivalent multimode bus waveguide bend with dual-core**. (a) The waveguide structure consisting of an input SWG, an equivalent multimode waveguide bend with a given bending radius $R$, and an output SWG; Insets: cross sections of the equivalent dual-core optical waveguide and the input/output SWGs; The calculated excess losses (b-d, respectively) and inter-mode crosstalk (e-g, respectively) for the cases of $\eta_0$ = 0.5, 0.4, and 0.6 as the bending radius $R$ and the duty-cycles ($\eta_1$, $\eta_2$) varies.

As an example, the core width is chosen as $w_{co}$=1.2μm to support three guided modes, and the etching depth of meta-surfaced layer is chosen to be $h_e$=80nm. Figure 2(b)-(g) show the calculation results for the excess losses and the inter-mode crosstalk when any one of three guided-modes in the input SWG is launched at the input port. Here the center duty-cycle $\eta_0$ is assumed to be 0.4, 0.5, or 0.6, while the edge duty-cycles $\eta_1$ and $\eta_2$ are given as $\eta_1= \eta_0 + \gamma\, w_{co}/2$ and $\eta_2= \eta_0 - \gamma\, w_{co}/2$, respectively. As the constant $\gamma$ can be chosen in the range of $0\leq \gamma \leq$**min**$[\eta_0/(w_{co}/2), (1-\eta_0)/(w_{co}/2)]$, one has ($0.4 \leq \eta_1 \leq 0.8$, $0 \leq \eta_2 \leq 0.4$), ($0.5 \leq \eta_1 \leq 1.0$, $0 \leq \eta_2 \leq 0.5$), and ($0.6 \leq \eta_1 \leq 1.0$, $0.2 \leq \eta_2 \leq 0.6$) for the cases of $\eta_0$=0.4, 0.5, and 0.6, respectively.

From Figure 2(b)-(d), it is apparent that that there are optimal values of the edge-duty-cycles ($\eta_{10}$,

$\eta_{20}$) for minimizing the excess loss as well as the inter-mode crosstalk when the center duty-cycle $\eta_0$ and the bending radius $R$ are given. In order to enable a smaller bending radius $R$, one should introduce a meta-surfaced structure with more asymmetry by choosing a larger optimal value $\eta_{10}$ and a smaller optimal value $\eta_{20}$, as shown in Figure 2(b)-(d). For example, for the case with $\eta_0$=0.5 [see Figure 2(b) and (e)], the optimal values ($\eta_{10}$, $\eta_{20}$) is modified from (0.7, 0.3) to (0.9, 0.1) when reducing the bending radius $R$ from 20μm to 10μm. The excess loss and the crosstalk for the bend designed with the optimal values ($\eta_{10}$, $\eta_{20}$) increases slightly as the bending radius $R$ decreases because the compensation is more difficult. Nevertheless, it is still possible to achieve very low excess loss (<0.4 dB) and very low crosstalk (<−30 dB) even when the bending radius is very small. For example, for a sharp bend with $R$=10μm, the excess loss is as low as 0.23 dB and the crosstalk is as low as −32 dB when choosing $\eta_0$=0.5 [see Figure 2(b) and (e)]. This shows that the mode mismatch between the BWG and the SWG can be compensated well by introducing asymmetric meta-surface structure.

It can also be seen that one can obtain similar results when choosing different duty-cycles $\eta_0$ ($\eta_0 \neq 0.5$), e.g., $\eta_0$=0.4 and 0.6. When choosing $\eta_0$=0.4 and 0.6, the corresponding optimal values ($\eta_{10}$, $\eta_{20}$) for a sharp bend with $R$=10μm are (0.8, 0) and (1, 0.2), respectively, as shown in Figure 2(b), 2(c), 2(f), and 2(g). When $\eta_0$=0.4, one has the design with $\eta_{20}$=0, which means that the width of the gaps at the outer edge is zero. When $\eta_0$=0.6, one has the design with $\eta_{10}$=1, which means that the width of the tips at the inner edge is zero. As a result, the fabrication becomes very difficult. Therefore, in this paper we focus on the optimal design for the MMWB with $\eta_0$=0.5.

Figure 3(a)-(c) show the simulated light propagation in the silicon photonic circuit consisting of input/output SWGs as well as the designed 90º-MMWB with $R$=10 μm and $w$=1.2 μm. Here the TE$_0$, TE$_1$ and TE$_2$ modes of the SWG are launched at the input port, respectively. The other structure parameters are given as $p$ = 250 nm, $\eta_0$ = 0.5, $\eta_1$ = 0.9, $\eta_2$ = 0.1, $h_{Si}$ = 220 nm, and $h_e$ = 80 nm. It can be seen that all these three modes propagate along the SWG-MMWB-SWG with low excess loss and low inter-mode crosstalk. The mode-field profiles at the input/output ends, as shown in the insets, are well matched with the straight waveguide modes. We also calculate the transmissions $\xi_{ij}$ (i.e., the mode excitation ratios) from the $i$-th guided-mode (TE$_i$) launched at the input SWG to the $j$-th guided-mode (TE$_j$) at the output SWG by using the mode expansion method, as shown in Figure 3(g)-(i). From the calculated $\xi_{ii}$ ($i$=1, 2, 3), one can estimate the excess loss for the $i$-th guided-mode propagation. Figure 3(d)-(f) show that the excess loss for the TE$_0$, TE$_1$ and TE$_2$ modes are respectively less than 0.3 dB, 0.2 dB and 0.5 dB in the broad band from 1500 nm to 1600 nm. Meanwhile, the inter-mode crosstalk is very low (<−30 dB) for all the guided modes. We also check the propagation of three guided modes along the silicon photonic circuit consisting of an input SWG, an S-bend with two cascaded 90º-MMWB, and an output SWG. The calculated results are shown in Figure 3(g)-(i). It can be seen that the excess losses and the inter-mode crosstalk increase slightly, as verified by the experimental results shown below.

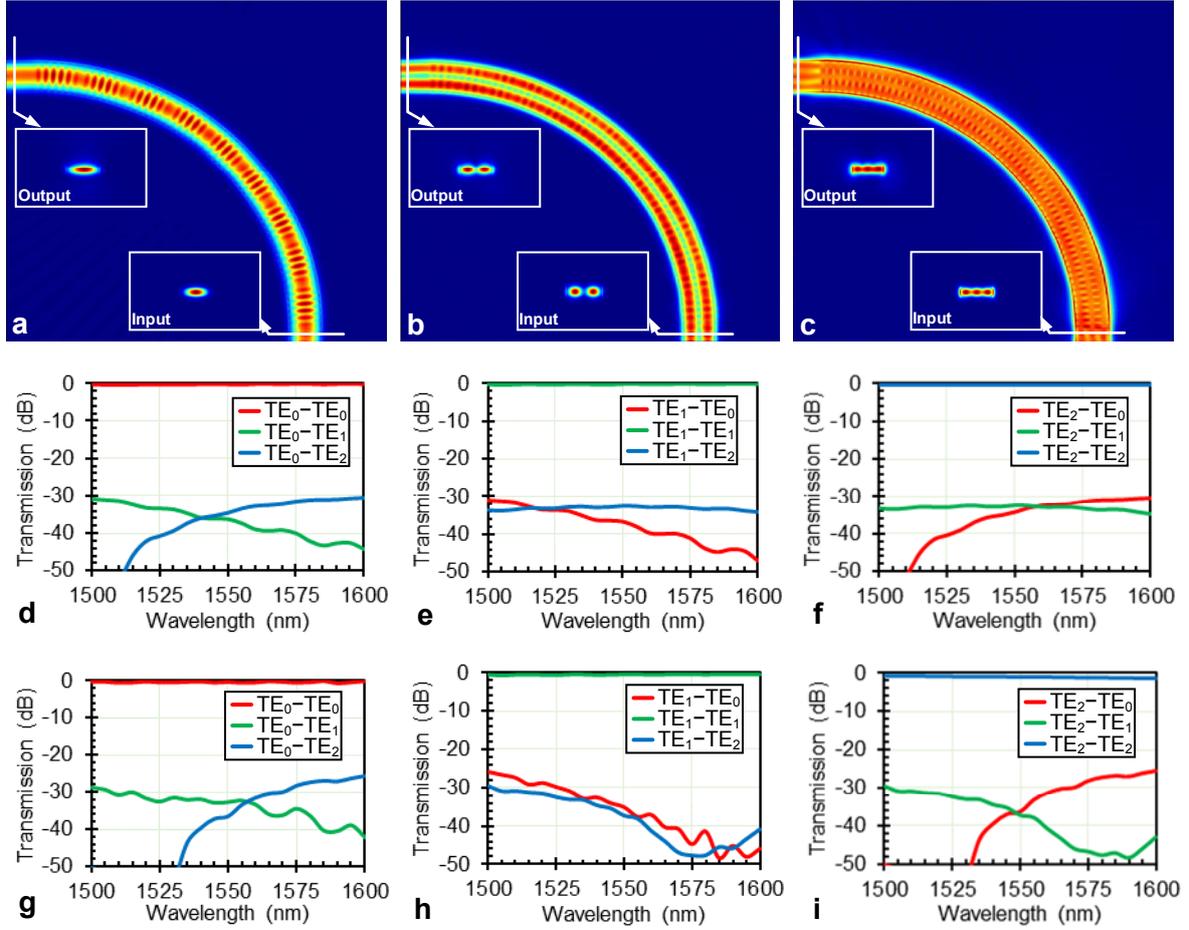

**Figure 3: Numerical simulation for the designed MMWB with three modes.** Figures (a-c) respectively show the simulated light propagation in the silicon PIC consisting of the input/output SWGs and the designed 90º-MMWB when different guided-modes ($TE_0$~$TE_2$) are launched from the input SWG, respectively. The wavelength dependence of the transmissions is also calculated for a silicon PIC consisting of a 90º-MMWB (d-f) or an S-bend with two cascaded 90º-MMWB (g-i) when the $TE_0$, $TE_1$ and $TE_2$ modes of the input SWG are launched, respectively. Here the MMWB is designed with the following parameters $R$=10 μm and $w$=1.2 μm.

The proposed approach for MMWB can be extended for more than three mode-channels. For example, we consider the design with $w_{co}$=1.8 μm for supporting *four* guided modes of TE polarization. In this case, the bending radius is marginally increased to $R$=20 μm for achieving low excess losses and low inter-mode crosstalk. The other parameters for the MMWB designed optimally are given as $p$ = 250 nm, $\eta_0$ = 0.5, $\eta_1$ = 0.9, $\eta_2$ = 0.1, $h_{Si}$ = 220 nm, and $h_e$ = 80 nm. Figure 4(a)-(d) show the simulated light propagation in the designed MMWB when the $TE_i$ mode ($i$=1, 2, 3, 4) is launched from the input SWG, respectively. The optical field profiles at the input/output ends are also shown in the insets. It can be seen that all four modes propagate along the SWG-MMWB-SWG with low excess losses and low inter-mode crosstalk. We also calculate the transmissions $\xi_{ij}$ from the $i$-th guided-mode ($TE_i$) launched at the input SWG to the $j$-th guided-mode ($TE_j$) at the output SWG, as shown in Figure 4(e)-(h). The calculations show that the excess loss for the $TE_0$, $TE_1$, $TE_2$ and $TE_3$ modes are 0.2~0.45 dB while the inter-mode crosstalk are lower than −24 dB in the broad band from 1500 nm to 1600 nm. Similarly, we also check the propagation of four guided-modes along the silicon PIC consisting of an input SWG, an S-bend with two cascaded 90º-MMWBs, and an output SWG. The calculated results are shown in Figure 4(i)-(l). It can be seen that the excess losses and the inter-mode crosstalk increase in comparison with the case with a single 90º-MMWB.

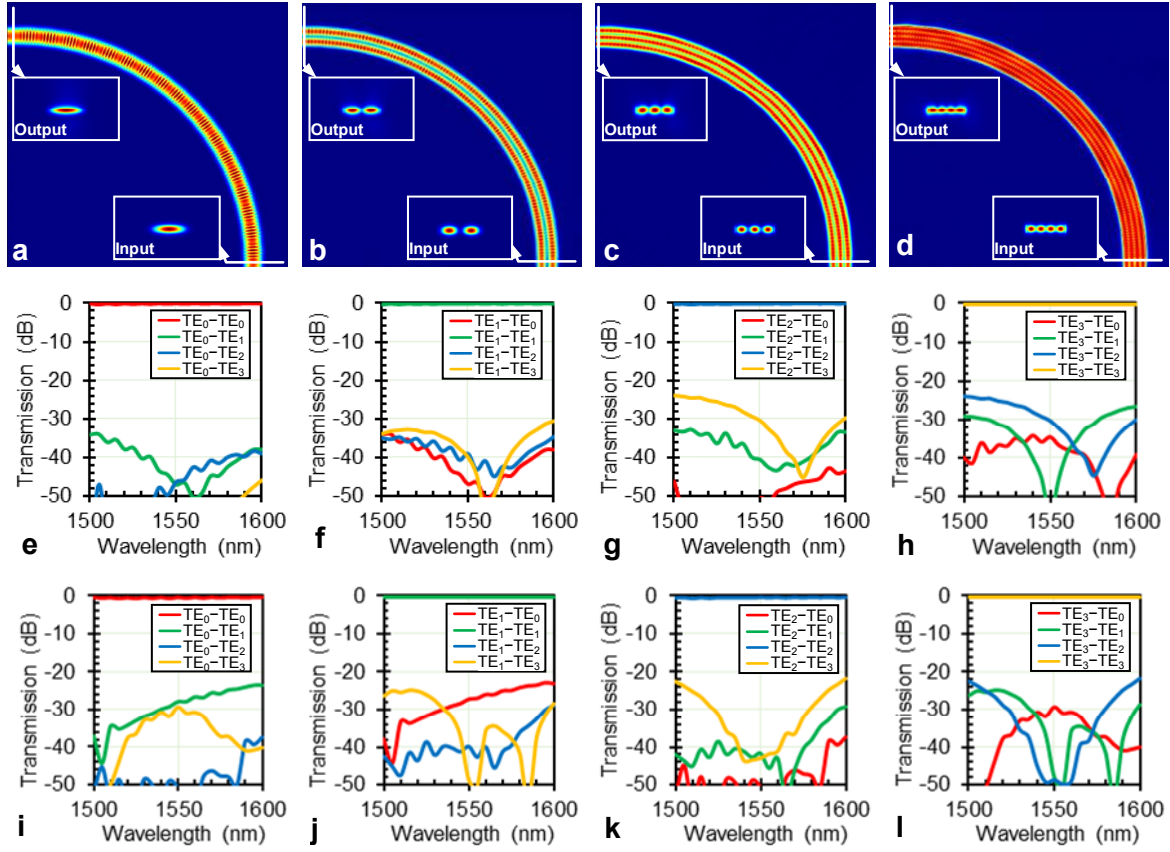

**Figure 4: Numerical simulation for the designed MMWB with four modes.** Figures (a-d) respectively show the simulated light propagation in the silicon PIC consisting of the input/output SWGs and the designed 90º-MMWB when different guided-modes ($TE_0 \sim TE_3$) are launched from the input SWG, respectively. The wavelength-dependences of the transmissions are also calculated for a silicon PIC consisting of a 90º-MMWB (e-h) or an S-bend with two cascaded 90º-MMWBs (i-l) when the $TE_0$, $TE_1$, $TE_2$ and $TE_3$ modes of the input SWG are launched, respectively. Here the MMWB is designed with the following parameters $R$=20 μm and $w$=1.8 μm.

These two MMWBs designed for supporting three and four guided-modes are then fabricated with nanofabrication technologies (see Device Fabrication). Figure 5(a) shows the microscope picture of the fabricated silicon PICs consisting of the designed MMWBs. Here S-bends with two cascaded 90°-MMWBs were introduced so that it is convenient to be measured with vertical fiber probes. An S-bend consists of two cascaded 90°-MMWBs, as shown in Figure 5(b), and Figure 5(c) shows the scanning electron microscope (SEM) pictures of the 90°-MMWBs. The mode (de)multiplexers with three or four mode-channels are connected at the input/output ends so that the transmission of any one mode-channels can be characterized selectively by launching the light from any selected input/output port. We also fabricated the straight multimode bus waveguide connected with the mode (de)multiplexers as the reference for the characterization of the present MMWBs.

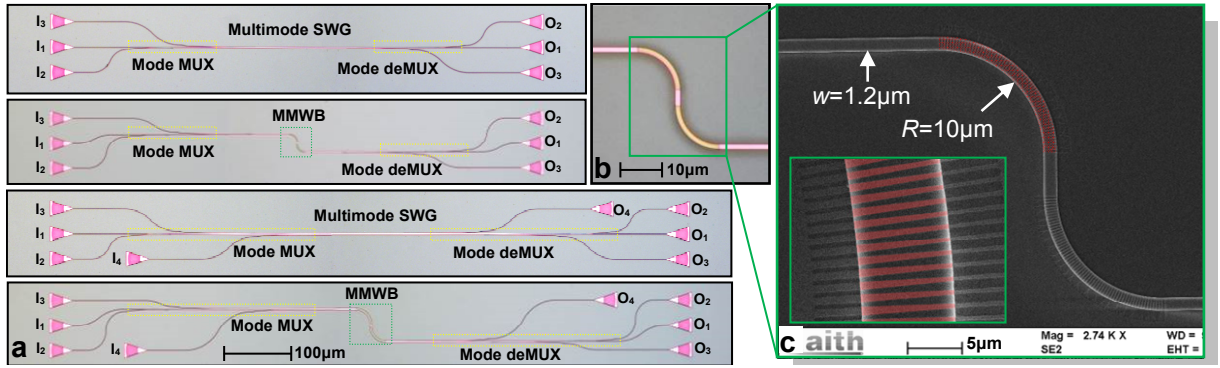

**Figure 5: Fabricated silicon PICs.** (a) Microscope pictures of the silicon PICs consisting of mode (de)multiplexers and a multimode bus waveguide with/without the S-bend (i.e., cascaded 90°-MMWBs); (b) Microscope picture of the fabricated S-bend with two cascaded 90°-MMWBs; (c) SEM picture of the fabricated S-bend with two cascaded 90°-MMWBs (inset: the meta-surface structure).

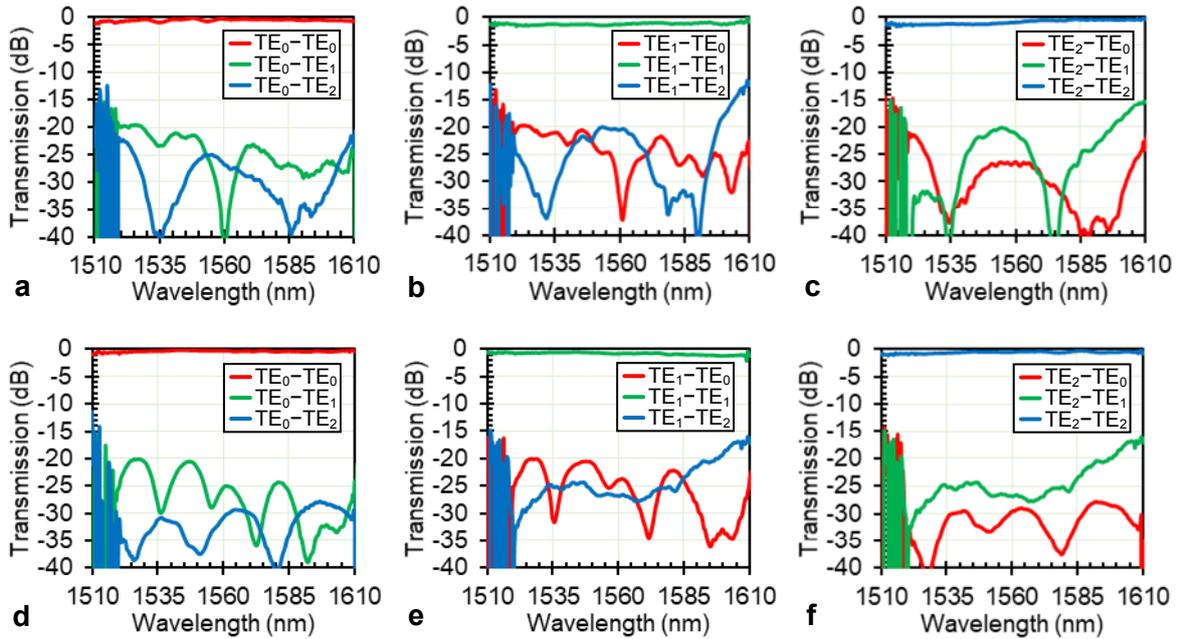

**Figure 6: Measurement results for the silicon PICs with three mode-channels. It consists of mode (de)multiplexers and a multimode bus waveguide with/without the S-bend (i.e., two cascaded 90°-MMWBs).** Measured spectral responses at the output ports ($O_1$, $O_2$, $O_3$) when one of the mode-channels (i.e., $TE_0$, $TE_1$, $TE_3$) is launched from the corresponding input port and goes through the multimode bus waveguide with (a-c) or without (d-f) an S-bend. Here $R$=10 μm and $w$=1.2 μm.

Figure 6(a)-(c) show the measured spectral response at the output ports ($O_1$, $O_2$, $O_3$,) when one of the mode-channels (i.e., $TE_0$, $TE_1$, $TE_3$) is launched from the corresponding input port and goes through the multimode bus waveguide consisting of an S-bend with $R$=10μm. The S-bend has two cascaded 90°-MMWBs. The transmissions of the three mode-channels in a straight multimode bus waveguide are also measured and shown in Figure 6(d)-(f), which can be used as a reference for evaluating the excess losses and inter-mode crosstalk introduced by the MMWBs. From Figure 6(d)-(f), it can be seen that the total excess losses are 0.5~1.1 dB and the inter-mode crosstalk is ~−22 dB in the wavelength range of 1520~1600 nm even when using a *straight* multimode bus waveguide. This is due to the undesired mode-coupling in the mode (de)multiplexers. In contrast, when introducing the S-bend with a sharp bending radius $R$=10μm as designed, the total excess losses are 0.9~1.9 dB and the inter-mode crosstalk is <−20 dB over a wavelength-band of >80 nm (1520~1600 nm). It can be seen that the measured excess losses and inter-mode crosstalk are very similar to those results shown in Figure 6(d)-(f). This indicates that the S-bend inserted does not introduce notable excess losses as well as inter-mode crosstalk, which is consistent with the simulation prediction shown in Figure 3(d)-(f). The excess loss of a single 90°-MMWG is estimated as 0.3~0.7dB by normalizing the loss of the mode (de)multiplexers. The increase in excess losses and crosstalk at short and long wavelength (<1520 nm or >1600 nm) is mainly due to the measurement error limited by the bandwidth of grating couplers.

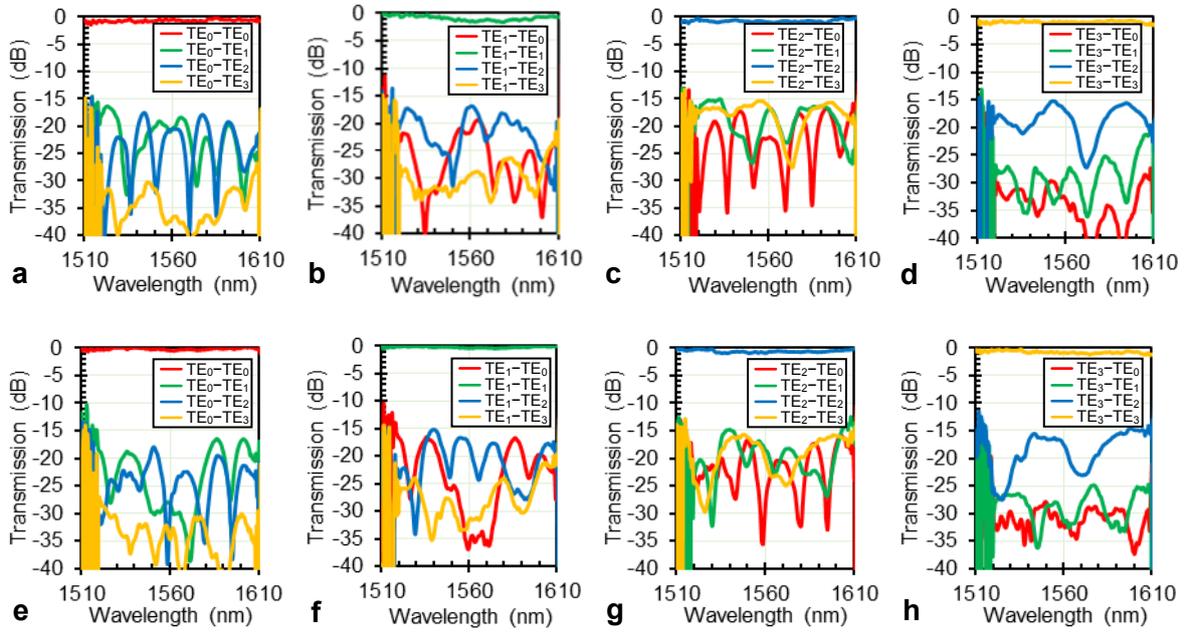

**Figure 7: Measurement results for the silicon PICs with four mode-channels. It consists of mode (de)multiplexers and a multimode bus waveguide with/without the S-bend (i.e., cascaded 90°-MMWBs).** Measured spectral response at the output ports ($O_1$, $O_2$, $O_3$, $O_4$) when one of the mode-channels (i.e., $TE_0$, $TE_1$, $TE_3$, $TE_4$) is launched from the corresponding input port and goes through the multimode bus waveguide with (a-d) or without an S-bend (e-h). Here $R$=20 μm and $w$=1.8 μm.

The fabricated silicon PIC consisting of the designed MMWB for four mode-channels was also characterized in a similar way. Figure 7(a)-(c) show the measured spectral responses at the output ports ($O_1$, $O_2$, $O_3$, $O_4$) when one of the mode-channels (i.e., $TE_0$, $TE_1$, $TE_2$, $TE_4$) launched from the corresponding input port goes through the multimode bus waveguide consisting of an S-bend based on two cascaded 90°-MMWBs with $R$=20 μm. The transmissions of these four mode-channels in a straight multimode bus waveguide were also measured and shown in Figure 7(d)-(f) as the reference.

By comparing these measurement results, one sees that the MMWB works well without introducing notable excess losses and inter-mode crosstalk for the four mode-channels. These measurement results show that a single 90°-MMWB has excess losses of ~0.5 dB, ~0.8 dB, ~0.5 dB and ~0.4 dB for the $TE_0$, $TE_1$, $TE_2$ and $TE_3$ modes, respectively, while the crosstalk is <−15 dB over a 90 nm (1520~1610 nm) wavelength-band.

# Discussion

Table 1 shows a comparison of our MMWB with other reported multimode waveguide bends on silicon. Compared to the structure proposed in [35], the MMWB may easily be extended to more mode-channels ($M_{ch} \geq 3$). The MMWB demonstrated in this paper has lower excess losses, lower inter-mode crosstalk, and sharper bending than the structure designed with graded height proposed in [35]. Furthermore, the MMWB can be fabricated with a regular lithography and dry-etching process. It can be seen that the present MMWB provides an excellent option to enable the low-loss and low-crosstalk propagation in multimode waveguide bends with multiple mode-channels.

Table 1: Summary of the reported multimode waveguide bends on silicon.

| Ref. | Structures | $M_{ch}$ | R (μm) | EL (dB) T | EL (dB) M | CT (dB) T | CT (dB) M | BW (nm) T | BW (nm) M |
|---|---|---|---|---|---|---|---|---|---|
| [33] | Mode converter | 2 | 5 | 0.12 | 0.2 | −23 | −22 | 100 | 100 |
| [35] | Gradational height | 3 | 78.8 | 2.5 | 2.6 | / | / | / | / |
| This work | MMWB | 3 | 10 | 0.2~0.5 | 0.3~0.7 | −30 | −20 | 100 | 80 |
| This work | MMWB | 4 | 20 | 0.2~0.5 | 0.4~0.8 | −24 | −15 | 100 | 90 |

Note: T-Theory; M-Measurement.

As a summary, this paper has proposed and demonstrated a novel MMWB enabling low-loss and low-crosstalk multi-channel transmission in a multimode bus waveguide with sharp bends. An asymmetric meta-surfaced structure has been introduced to compensate the structural asymmetry introduced by the sharp-bending, in which the mode mismatching between the straight section and the bent section is reduced significantly. The multimode waveguide bends with a bending radius of 10 μm and 20 μm have been realized for the cases with three and four mode-channels, respectively. For the case with three mode-channels, it has been shown in theory that the present ultra-sharp MMWB has a low excess loss (0.2~0.5 dB) and a low inter-mode crosstalk (<−30 dB) over a broad wavelength-band (>100 nm), which is verified experimentally by comparing the measured transmission of all the mode-channels in the silicon PICs with or without S-bends (i.e., two cascaded 90º-MMWB). It has also been verified that the present MMWB can be extended for the transmission with more mode-channels. The present work paves the way for using meta-surfaced silicon photonic waveguide structures for on-chip multimode manipulation.

# Methods

## Simulation

Lumerical MODE software with uniform grid sizes (10nm) were used for the numerical simulations, including the analyses of the mode fields in SWGs, BWGs, and MMWBs, as well as the calculation of the overlap integral between different modes.

The 3D finite-difference time-domain (FDTD) method provided by Lumerical FDTD was used to calculate the light propagation in the designed silicon PICs and the output optical fields with non-uniform grid sizes.

The wavelength dependence of the excess losses and the inter-mode crosstalks for all the mode-channels were calculated with the mode expansion method provided by Lumerical FDTD.

Assume that the $j$-th mode at the output SWG has a mode excitation ratio $\xi_{ij}$ when the $i$-th mode is launched at the input SWG and propagates along a given silicon PIC. The excess loss for the $i$-th mode-channel is given by

$$EL_i=\log_{10}(\xi_{ii}).$$

The inter-mode crosstalk from the $i$-th mode-channel to the $j$-th mode-channel is then given by ($j \neq i$)

$$CT_{ij}=\log_{10}(\xi_{ij}).$$

For the design of the present MMWB, the maximum of $CT_{ij}$ is evaluated for the optimization, i.e., $CT_{j,\max}=\max(CT_{1j}, CT_{2j}, \ldots, CT_{Ij})$.

## Fabrication

All devices were fabricated on a 220nm-thick silicon-on-insulator (SOI) wafer with a 2μm-thick $SiO_2$ under-cladding layer. An electron-beam lithography (EBL) process with the MA-N2403 photoresist and an inductively coupled plasma (ICP) dry-etching process were carried out to fabricate the straight/bent waveguides on the top-silicon layer. Another shallowly-etched overlay process with an etching step of 80nm was applied to fabricate the meta-surfaced structures in the bent section as well as the grating couplers for fiber-chip coupling. A $SiO_2$ upper-cladding was deposited by using plasma enhanced chemical vapor deposition (PECVD).

## Characterization

For the characterization of the fabricated chips, a broad-band amplified spontaneous emission (ASE) light source (1510~1610 nm) was used at the input side. The total optical power of the ASE is about 16 dBm. An optical spectrum analyzer (OSA) was used to receive light at the output side. The OSA has a high resolution (~0.02 nm). On-chip grating couplers at the input/output ends were used for achieving efficient fiber-chip coupling. The grating coupler has a coupling efficiency of ~9.1 dB (@ 1550nm) and a ~40 nm 3dB-bandwidth.

# Acknowledgements

This work was supported by National Natural Science Foundation of China (NSFC) (61725503, 11374263, 61422510, 61431166001), Zhejiang Provincial Natural Science Foundation (Z18F050002), and National Major Research and Development Program (No. 2016YFB0402502) and NSFC-RGC joint research scheme N_CUHK404/14.

# Author contributions

D. D. conceived the idea of MMWB. H. W. performed the theoretical analysis, design and characterization of the devices. C. L. did the fabrication, took the SEM and microscope images. L. S. helped the fabrication and characterization. H. W. and D. D. wrote the manuscript. H. W., C. L., L. S, H. T., J. B, and D. D. revised the manuscript and contributed to the discussions. D. D. supervised the project.

# Competing financial interests

The authors declare no competing financial interests.